\documentclass[twocolumn]{aastex62}
\usepackage{csvsimple}
\usepackage{amsmath}	
\usepackage{amssymb}	
\usepackage{multirow}

\newcommand{\kep}{{\it Kepler}}
\newcommand{\Rhk}{$\log\,R'_\text{HK}$}
\newcommand{\dnu}{$\delta\nu$}
\newcommand{\numax}{$\nu_\text{max}$}
\newcommand{\prot}{$P_\text{rot}$}
\newcommand{\sph}{$S_\text{ph}$}
\newcommand{\teff}{$T_\text{eff}$}
\newcommand{\metal}{$\text{[Fe/H]}$}
\newcommand\aastex{AAS\TeX}

\shorttitle{\aastex\ Magnetic activity seen through asteroseismology}
\shortauthors{A. R. G. Santos et al.}

\usepackage{hyperref}

\begin{document}

\title{
\MakeUppercase{Signatures of magnetic activity:}

\MakeUppercase{On the relation between stellar properties and p-mode frequency variations}}

\author{A. R. G. Santos}
\email{asantos@spacescience.org}
\affil{Space Science Institute, 4750 Walnut Street, Suite 205, Boulder CO 80301, USA}

\author{T. L. Campante}
\affil{Instituto de Astrof\'{i}sica e Ci\^{e}ncias do Espa\c{c}o, Universidade do Porto, CAUP, Rua das Estrelas, PT-4150-762 Porto, Portugal}
\affil{Departamento de F\'{i}sica e Astronomia, Faculdade de Ci\^{e}ncias, Universidade do Porto, Rua do Campo Alegre 687, PT-4169-007 Porto, Portugal}

\author{W. J. Chaplin}
\affil{School of Physics and Astronomy, University of Birmingham, Edgbaston, Birmingham B15 2TT, UK}
\affil{Stellar Astrophysics Centre, Department of Physics and Astronomy, Aarhus University, Ny Munkegade 120, DK-8000 Aarhus C, Denmark}

\author{M. S. Cunha}
\affil{Instituto de Astrof\'{i}sica e Ci\^{e}ncias do Espa\c{c}o, Universidade do Porto, CAUP, Rua das Estrelas, PT-4150-762 Porto, Portugal}
\affil{Departamento de F\'{i}sica e Astronomia, Faculdade de Ci\^{e}ncias, Universidade do Porto, Rua do Campo Alegre 687, PT-4169-007 Porto, Portugal}

\author{J. L. van Saders}
\affil{Institute for Astronomy, University of Hawai'i, 2680 Woodlawn Drive, Honolulu, HI 96822, USA}

\author{C. Karoff}
\affil{Department of Geoscience, Aarhus University, H\o egh-Guldbergs Gade 2, 8000, Aarhus C, Denmark}

\author{T. S. Metcalfe}
\affil{Space Science Institute, 4750 Walnut Street, Suite 205, Boulder CO 80301, USA}
\affil{Max-Planck-Institut f\"ur Sonnensystemforschung, Justus-von-Liebig-Weg 3, 37077, G\"ottingen, Germany}

\author{S. Mathur}
\affil{Instituto de Astrofisica de Canarias, Spain}
\affil{Universidad de La Laguna, Spain}

\author{R. A. Garc\'{i}a}
\affil{IRFU, CEA, Universit\'e Paris-Saclay, F-91191 Gif-sur-Yvette, France}
\affil{Universit\'e Paris Diderot, AIM, Sorbonne Paris Cit\'e, CEA, CNRS, F-91191 Gif-sur-Yvette, France}

\author{M. N. Lund}
\affil{Stellar Astrophysics Centre, Department of Physics and Astronomy, Aarhus University, Ny Munkegade 120, DK-8000 Aarhus C, Denmark}
\affil{School of Physics and Astronomy, University of Birmingham, Edgbaston, Birmingham B15 2TT, UK}

\author{R. Kiefer}
\affil{Centre for Fusion, Space, and Astrophysics, Department of Physics, University of Warwick, Coventry, CV4 7AL, UK}

\author{V. Silva Aguirre}
\affil{Stellar Astrophysics Centre, Department of Physics and Astronomy, Aarhus University, Ny Munkegade 120, DK-8000 Aarhus C, Denmark}

\author{G. R. Davies}
\affil{School of Physics and Astronomy, University of Birmingham, Edgbaston, Birmingham B15 2TT, UK}
\affil{Stellar Astrophysics Centre, Department of Physics and Astronomy, Aarhus University, Ny Munkegade 120, DK-8000 Aarhus C, Denmark}

\author{R. Howe}
\affil{School of Physics and Astronomy, University of Birmingham, Edgbaston, Birmingham B15 2TT, UK}
\affil{Stellar Astrophysics Centre, Department of Physics and Astronomy, Aarhus University, Ny Munkegade 120, DK-8000 Aarhus C, Denmark}

\author{Y. Elsworth}
\affil{School of Physics and Astronomy, University of Birmingham, Edgbaston, Birmingham B15 2TT, UK}
\affil{Stellar Astrophysics Centre, Department of Physics and Astronomy, Aarhus University, Ny Munkegade 120, DK-8000 Aarhus C, Denmark}

\begin{abstract}
In the Sun, the properties of acoustic modes are sensitive to changes in the magnetic activity. In particular, mode frequencies are observed to increase with increasing activity level. Thanks to CoRoT and \kep, such variations have been found in other solar-type stars and encode information on the activity-related changes in their interiors. Thus, the unprecedented long-term \kep\ photometric observations provide a unique opportunity to study stellar activity through asteroseismology. The goal of this work is to investigate the dependencies of the observed mode frequency variations on the stellar parameters and whether those are consistent with an activity-related origin.  We select the solar-type oscillators with highest signal-to-noise ratio, in total 75 targets. Using the temporal frequency variations determined in \citet{Santos2018}, we study the relation between those variations and the fundamental stellar properties. We also compare the observed frequency shifts with chromospheric and photometric activity indexes, which are only available for a subset of the sample. We find that frequency shifts increase with increasing chromospheric activity, which is consistent with an activity-related origin of the observed frequency shifts. Frequency shifts are also found to increase with effective temperature, which is in agreement with the theoretical predictions for the activity-related frequency shifts by \citet{Metcalfe2007}. Frequency shifts are largest for fast rotating and young stars, which is consistent with those being more active than slower rotators and older stars. Finally, we find evidence for frequency shifts increasing with stellar metallicity.
\end{abstract}

\keywords{asteroseismology -- stars: solar-type -- stars: oscillations -- stars: activity}

\section{Introduction} \label{sec:intro}

In solar-type oscillators, near-surface convection stochastically excites acoustic oscillations \citep[e.g.][]{Goldreich1977}. Convection is also a key ingredient for the magnetic field generation and, in particular, for the activity cycles \citep[e.g.][]{Brun2017}.

As the magnetic activity level changes in the Sun, reflected in the various activity proxies, such as 10.7-cm flux, sunspot number and area, photometric activity proxy, and chromospheric activity index, the properties of the acoustic modes are observed to change: mode frequencies increase with increasing activity while mode amplitudes decrease \citep[e.g.][]{Woodard1985,Elsworth1990,Libbrecht1990a,Chaplin1998,Howe2015,Salabert2017}.

The first evidence for activity-related frequency shifts in a star other than the Sun was found by \citet{Garcia2010}. The authors found the mode parameters of HD~49933 \citep[observed by CoRot - Convection, Rotation, and planetary Transits;][]{Baglin2006} varying over time, while its starspot proxy is also consistent with an activity cycle. Making use of the long-term \kep\ short-cadence data \citep{Borucki2010}, temporal frequency shifts, possibly activity-related, have been measured for several solar-type stars \citep{Salabert2016,Salabert2018,Regulo2016,Kiefer2017,Santos2018}. In particular, \citet{Salabert2016} found the frequency shifts varying approximately in phase with the photometric activity index \citep{Mathur2014} for a young solar analog KIC~10644253. Using spectroscopic data, the authors confirmed that KIC~10644253 is more active than the Sun. Another well studied case is KIC~8006161 \citep{Karoff2018}, which is another solar analog but with higher metallicity than the Sun. Using spectroscopic and photometric observations, and a number of different techniques and diagnostics for magnetic activity (including seismic diagnostics), KIC~8006161 was found to have significantly stronger differential rotation, both in the radial and latitudinal direction, and stronger activity cycle than the Sun. In particular, the amplitude of the activity cycle of KIC~8006161 seen through the chromospheric activity index is about 2.7 larger than that of the Sun. \citet{Karoff2018} interpreted these characteristics as the result of a deep convection zone due to the high metallicity of KIC~8006161. This suggests that metallicity may affect activity and thus the observed frequency shifts.

Besides the possible effect from metallicity addressed by \citet{Karoff2018}, frequency shifts may also depend on the stellar effective temperature. Assuming that frequency shifts scale linearly with the chromospheric activity, as observed over the solar cycle, \citet{Chaplin2007a} predicted that the activity-related frequency shifts decrease with effective temperature. By contrast, the theoretical predictions by \citet[][see \citet{Dziembowski2007} for a generalization to non-radial modes]{Metcalfe2007} are consistent with frequency shifts increasing with increasing effective temperature. In this formulation, frequency shifts depend on the chromospheric activity index, mode inertia, and on the depth of the source of the perturbations beneath the photosphere, which in turn is related to the pressure scaleheight at the photosphere. Both formulations \citep{Chaplin2007a,Metcalfe2007} predict frequency shifts decreasing with age, which is consistent with stars becoming less active as they evolve \citep[e.g.][]{Wilson1963,Wilson1964}. Also, as stars evolve the surface rotation rate decreases, often being used as a diagnostic for stellar ages in gyrochronology \citep[e.g.][]{Skumanich1972,Barnes2007,Garcia2014,Davies2015}. Thus, in general, faster rotators are found to be more active than slower rotators \citep[e.g.][]{Vaughan1981,Baliunas1983,Noyes1984b}. A tighter relationship is found between chromospheric activity and the Rossby number (ratio between rotation period and convective turnover timescale) with magnetic activity increasing as Rossby number decreases \citep[e.g.][]{Noyes1984b,Mamajek2008}.

In this work, we study the relation between the observed frequency shifts and stellar properties, namely effective temperature, rotation period, Rossby number, age, and metallicity. We use the frequency shifts obtained by \citet{Santos2018}, who analyzed 87 oscillating solar-type stars observed by \kep. The authors measured frequency shifts from 90-day segments of the original time-series. The individual mode frequencies were obtained through a Bayesian global peak-bagging analysis, but only the five central orders (closest to the frequency of maximum power, \numax, with the highest signal-to-noise ratio) were used to compute frequency shifts. The authors found that more than half of the targets show evidence for periodic variations in the mode frequencies. 

For stars with available chromospheric and photometric activity measurements, we also compare the observed frequency shifts with the activity proxies \Rhk\ \citep{Noyes1984b} and \sph \citep{Mathur2014}.

\section{Target sample}\label{sec:sample}

We start by taking the target sample considered in \citet{Santos2018}, which comprises 87 \kep\ solar-type stars including the LEGACY sample \citep{Lund2017,SilvaAguirre2017} and 21 additional \kep\ Objects of Interest \citep[KOIs, e.g.][]{SilvaAguirre2015,Davies2016}.

\citet{Santos2018} measured temporal variations in the mode frequencies of the 87 \kep\ solar-type stars. However, some of the targets, namely some KOIs, have very low signal-to-noise ratio, which hampers the task of constraining the mode frequencies from subseries of length 90 days. The frequency variations of such targets are likely to be strongly affected by noise. Since the goal is to study activity-related frequency shifts, we remove the targets with lowest signal-to-noise from the current analysis. In the next section we take further steps to ensure that the frequency variations are not noise.

\begin{figure}[h]
\includegraphics[width=\hsize]{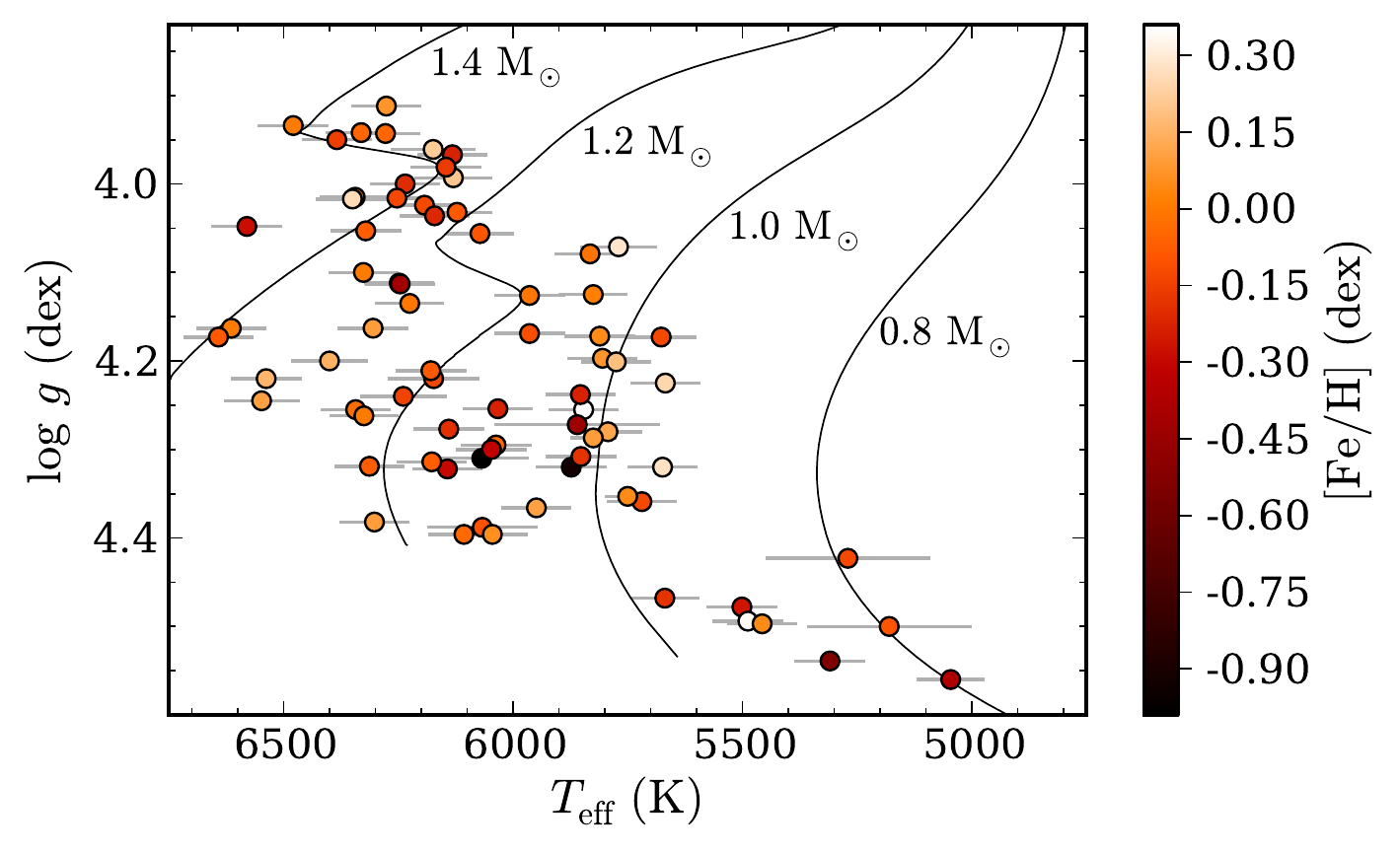}
\caption{Kiel-diagram for the target sample color coded by metallicity. The black solid lines show the solar-calibrated evolutionary tracks obtained with the evolution code Modules for Experiments in Stellar Astrophysics \citep[MESA;][]{Paxton2011,Paxton2013}.}\label{fig:HR}
\end{figure}

The signal-to-noise ratio for a given subseries is measured as the average ratio between the mode heights for the five central orders and the background signal \citep[both from the analysis in][]{Santos2018}. The final signal-to-noise ratio for each target corresponds to the average value over all the 90-d subseries with a duty cycle greater than $70\%$. The targets with a signal-to-noise ratio lower than one are removed from the subsequent analysis: KIC~10730618 and ten KOIs (KIC~3425851, KIC~4141376, KIC~4349452, KIC~4914423, KIC~5866724, KIC~7670943, KIC~8478994, KIC~8494142, KIC~11401755, and KIC~11904151). We further remove the star KIC~3427720, whose light-curve shows sudden variations also visible in the properties measured in \citet{Santos2018}, in particular in the frequency shifts. The final target sample consisting of 75 stars is shown in Figure~\ref{fig:HR}.

In what follows, we study the relation between the observed frequency shifts and stellar properties. The values for effective temperature \teff, surface gravity $\log\,g$, metallicity \metal, and rotation period \prot, and the respective sources are listed in \citet{Santos2018}. We  note however that for KIC~7970740 we consider the rotation period \prot$\simeq42\pm7$ days (from Santos et al. submitted). The rotation period in the literature corresponds roughly to half of the true period of that star. We adopt the stellar ages determined by \citet{SilvaAguirre2015,SilvaAguirre2017}. The photometric activity index \sph\ corresponds to the average value of the data shown in \citet{Santos2018}. Finally, the values for the chromospheric activity and sources are listed in Appendix \ref{app}.

\section{Maximum frequency variation}\label{sec:deltanu}

\begin{figure}[h!]
\includegraphics[width=\hsize]{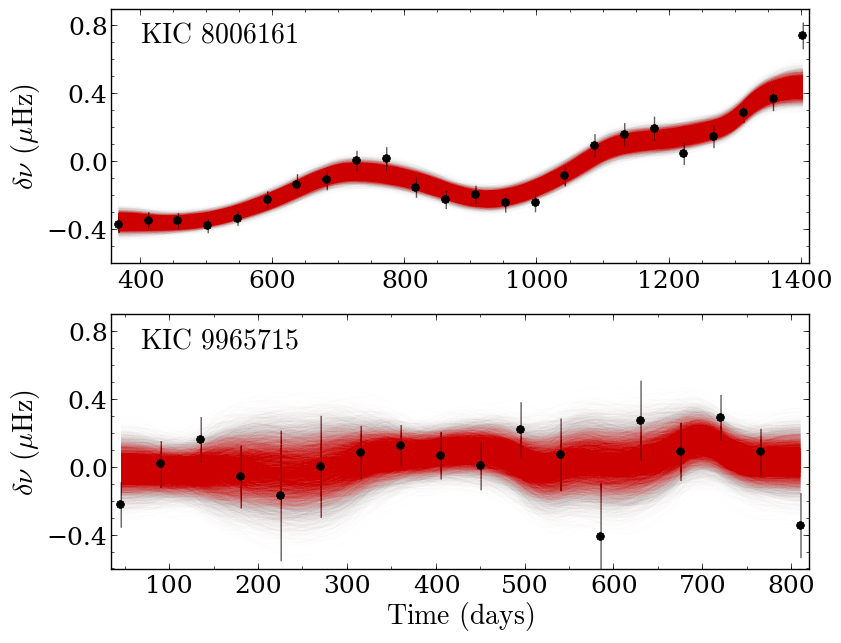}
{\caption{Frequency shifts are shown in black for KIC~8006161 (top) and KIC~9965715 (bottom). The transparent red lines represent $10^4$ individual realizations of the filtering procedure (see text). For illustration purposes the data points for the $10^4$ realizations are interpolated.}\label{fig:deltanu}}
\end{figure}

Throughout the target sample, frequency shifts, \dnu, exhibit different types of behavior showing evidence for: possible rising or declining phases of an activity cycle; more than one complete cycle; non-cyclic variations; or no variations. In order to be consistent, we decided to compute the amplitude of the frequency shifts based on the minimum and maximum value of the observed shifts. However, variations in the mode frequencies may be affected by instrumental noise or sudden changes in the quality of the data, and our inability to properly constrain the mode frequencies under those circumstances, which is also reflected in the error bars. Therefore, in order to prevent an overestimation of the intrinsic frequency variation, we follow a number of steps to compute the final maximum $\delta\nu$ variation:
\begin{enumerate}
	\item For each target, we perform $10^4$ realizations, in which \vspace{-0.2cm}
	\begin{enumerate}
		\item each \dnu\ data point randomly varies within the error bar according to a Gaussian distribution with mean and standard deviation being the frequency shift and respective uncertainty estimated by \citet{Santos2018};
		\item the data points are smoothed/filtered over 180 days;
		\item we take the difference between the maximum and minimum value of the smoothed data.\vspace{-0.2cm}
	\end{enumerate}
	\item For each target, the maximum \dnu\ variation and uncertainty correspond to the median and standard deviation of the results for $10^4$ realizations.
\end{enumerate}
Figure \ref{fig:deltanu} shows an example of a relatively well behaved case (KIC~8006161) and an example of a case with larger uncertainties on the frequency shifts and sporadic variations (KIC~9965715), which may be consistent with no variation. In the case of KIC~8006161 we may underestimate the true variation in the frequency shifts, but for KIC~9965715 (and for other targets in the sample) we may overestimate the intrinsic variation by just taking the minimum and maximum values of the observed frequency shifts. Therefore, we opt for this more conservative approach.

For comparison, we also estimate the maximum frequency-shift variations for the Sun. We use the temporal frequency shifts measured for solar cycle 23 by \citet{Santos2018} from VIRGO/SPM data \citep[Variability of solar IRradiance and Gravity Oscillations on board SOlar and Heliospheric Observatory (SOHO), where SPM stands for sunphotometers;][]{Frohlich1995,Frohlich1997,Jimenez2002}. We measure the maximum \dnu\ variation computed from the complete 11-year cycle following the steps listed before. We also compute the range of possible values for the maximum \dnu\ variation when considering only four years of VIRGO/SPM data to be consistent with \kep\ observations. We take 4-year subseries spaced by one year, corresponding to different phases of the solar activity cycle. The range of possible values is determined by the minimum and maximum values found for the 4-year segments.

\section{Relation between frequency shifts and stellar properties}\label{sec:results}

Activity-related frequency shifts are expected to be common among solar-type oscillators. More than half of the solar-type stars analyzed by \citet{Santos2018} show significant temporal variations in the frequencies of the acoustic modes, often accompanied by variations in other parameters. However, to properly connect those variations to an activity-related origin, more data are needed (for example spectroscopy).

\begin{figure*}[h!]\centering
\includegraphics[width=1.\hsize]{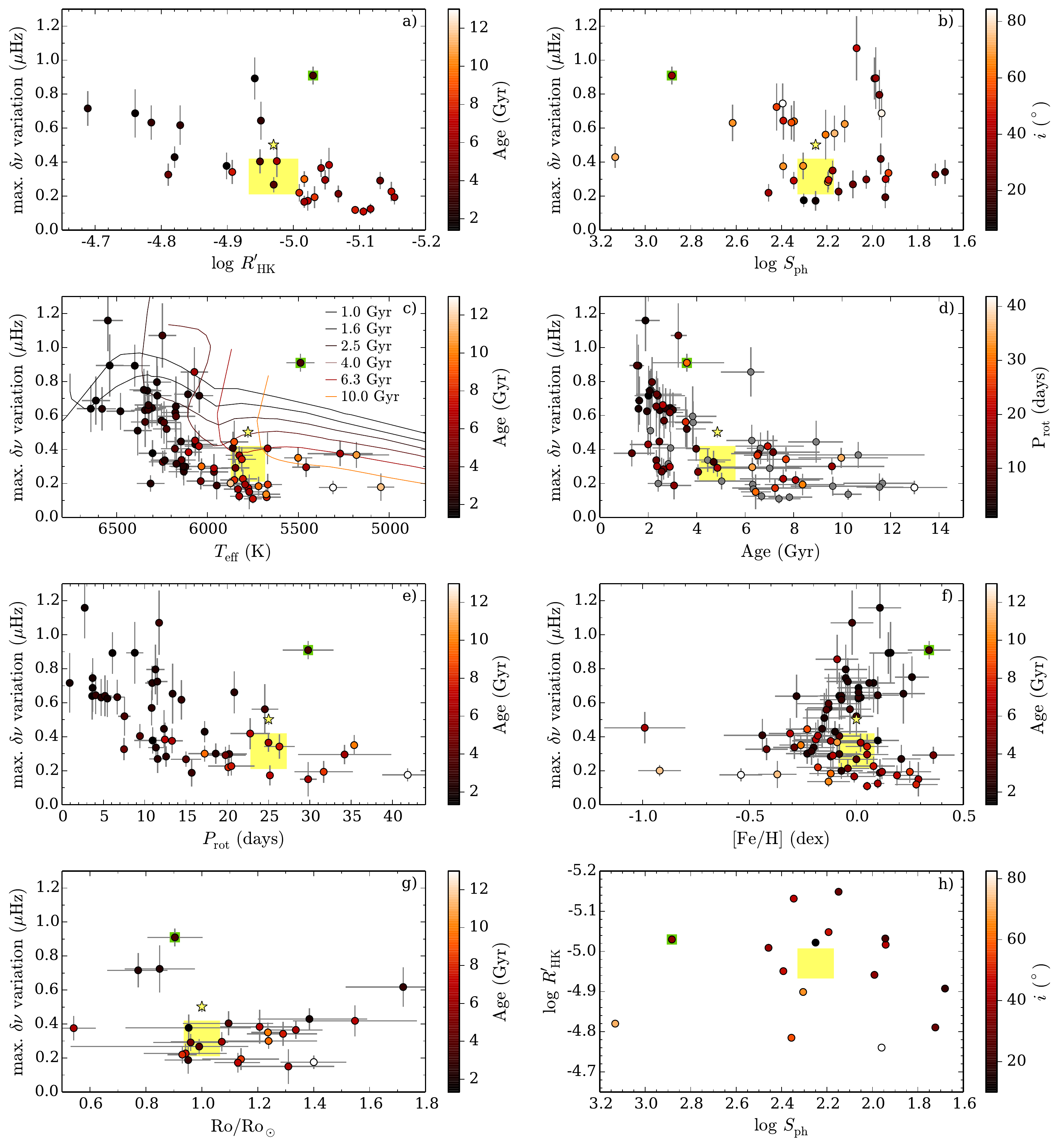}
\caption{Maximum \dnu\ variation as a function of: a) chromospheric activity index, \Rhk; b) photometric activity proxy, \sph; c) effective temperature, \teff; d) age; e) rotation period, \prot; f) metallicity, \metal; and g) Rossby number, Ro. Panel~h) shows the relation between the chromospheric and photometric activity indexes. All panels are color coded by age, except panels b) and h), and d) which are color coded by inclination angle and \prot\ (stars with unknown \prot\ in gray), respectively. Solid lines in panel c) show the theoretical prediction of activity-related \dnu\ for different ages by \citet[][adapted from \citet{Karoff2009}]{Metcalfe2007}. Yellow star marks the maximum frequency shift variation for the complete solar cycle 23. Yellow shaded region indicates the range of values from 4-year subseries of the solar cycle 23. Its width is set at $10\%$ of the parameter range shown. KIC~8006161 is highlighted by the green square.}\label{fig:params}
\end{figure*}

Currently there are chromospheric activity measurements only for 30 of the 75 solar-type stars in the target sample and not all are contemporaneous to \kep\ observations. Panel a) of Figure~\ref{fig:params} shows the maximum frequency-shift variation as a function of the chromospheric activity index, \Rhk. Large frequency-shift values tend to correspond to high chromospheric activity level. We note however that for some stars the value for the maximum frequency-shift variation may be an underestimate of the total variation over a complete activity cycle. This is shown for the Sun through the difference between the yellow star (maximum \dnu\ variation for solar cycle 23) and the yellow shaded region (range of possible values measured from 4-year subseries of the solar cycle~23). Furthermore, for some targets, the chromospheric and photometric observations are not contemporaneous (see Appendix \ref{app} and references therein), which may contribute to additional scatter. The Spearman's correlation coefficient is $\sim0.69$ which indicates a strong correlation between \dnu\ and \Rhk. Panel b) of Figure~\ref{fig:params} shows the frequency shifts as a function of the photometric activity index \sph, available for 34 stars. Based on a sample of solar analogs, \citet{Salabert2016a} showed that the photometric and the chromospheric activity indexes are complementary. However, for our target sample, no clear relation is found between frequency shifts and \sph (Spearman's correlation coefficient of $\sim0.13$). For some stars, the value of \sph\ is a lower limit on the true activity, as it depends on the visibility of active regions and, thus, on their latitudinal distribution and stellar inclination. The color code in this panel indicates the seismic inclination angle obtained by \citet{Lund2017}. Assuming the solar distribution of active regions, \sph\ is more likely to be underestimated for darker colored (smaller inclination) than for lighter colored targets. For comparison, panel~h) shows the relation between the chromospheric and photometric activity indexes for the targets with both measurements. The disagreement is greatest for low \sph\ values which, with exception of one target, correspond to stars with low inclination angle. Nevertheless, the comparison shown in panel a) of Figure~\ref{fig:params} suggests that the observed frequency shifts are well correlated to the chromospheric activity level, supporting the hypothesis of the measured frequency shifts being activity-related.

Next, we search for correlations between the amplitude of the observed variations in the frequencies of the acoustic modes and other stellar properties.

Panel c) of Figure \ref{fig:params} shows the maximum frequency-shift variation as a function of effective temperature for the 75 solar-type stars color coded by age. Frequency shifts are found to increase with effective temperature, with a Spearman's correlation coefficient of $\sim0.68$. This indicates a strong correlation between frequency shifts and effective temperature. Our results are qualitatively in agreement with the theoretical prediction by \citet{Metcalfe2007}, which for comparison is shown by the solid lines with the same color code as the data points. \citet{Chaplin2007a} predicted a decrease of the frequency shifts with effective temperature, which is not consistent with our results. There is a clear outlier in the relationship between \teff\ and \dnu\ at $\sim5488$ K, which shows larger frequency shifts than its neighbors. This outlier is the metal-rich star KIC~8006161, which was found to have a significantly stronger activity cycle than the Sun \citep[see][]{Karoff2018}.

As mode linewidths increase with effective temperature \citep[e.g.][]{Appourchaux2012,Lund2017}, at higher temperature, the mode frequencies are more poorly constrained. This is also reflected on the uncertainties. Thus, one may expect an increasing variability in the frequencies with effective temperature just as a result of noise. This could bring into question the origin of the observed frequency shifts. However, the procedure followed in Section~\ref{sec:deltanu} should reduce this noise contribution. Moreover, based on Monte Carlo simulations, \citet{Salabert2018} have selected a subsample of 20 stars for which the frequency shifts are less likely to be just noise. 17 of those stars are common to the sample studied in this work. Their effective temperatures are between 5500 K and 6600 K, and they follow the relation seen in panel c) of Figure~\ref{fig:params}. We are, therefore, confident that the frequency-shift dependence on effective temperature found here is real.

Panel c) of Figure~\ref{fig:params} also shows a dependency with age, where younger stars tend to have larger variations in the frequencies of the acoustic modes than older stars. That is also illustrated in panel d) of Figure~\ref{fig:params}. In the \dnu-age diagram, KIC~8006161 (orange symbol with $\text{Age}\sim3.59$ Gyr and maximum \dnu\ variation of $0.91\mu$Hz) is no longer isolated. The results are consistent with younger stars being more active than older stars \citep[e.g.][]{Wilson1963,Wilson1964}. The Spearman's correlation coefficient is $\sim-0.66$ confirming a strong negative correlation between frequency shifts and age.

Panel e) of Figure~\ref{fig:params} shows the frequency-shift variations as a function of the surface rotation period. In general, frequency shifts are larger for fast rotating stars, with KIC~8006161 ($P_\text{rot}\sim29.79$ d) being an outlier in this relation. Our results are consistent with the expectation of fast rotators being more active than slower rotators \citep[e.g.][]{Vaughan1981,Baliunas1983,Noyes1984b}. The Spearman's correlation coefficient is $\sim-0.61$ which agrees with a strong negative correlation between frequency shifts and rotation period.

Panel f) of Figure~\ref{fig:params} shows the frequency variation as a function of metallicity, color coded by age. There is no clear relationship between frequency shifts and metallicity (Spearman's correlation coefficient of 0.09). Nevertheless, there is evidence for two different sequences (see Appendix~\ref{app:cluster} for details). For younger stars the frequency shifts generally increasing with metallicity (Spearman's correlation coefficient of 0.39). This supports the interpretation by \citet{Karoff2018}, where the strong activity cycle was assigned to a deeper convection zone due the high metallicity of KIC~8006161 ($\text{[Fe/H]}=0.34$). For older stars, however, frequency shifts decrease with metallicity (Spearman's correlation coefficient of -0.46). Note that for some stars in the sample the value for maximum \dnu\ may be a lower limit of the activity-related frequency shifts as we only have access to part of the activity cycle.

Panel g) of Figure~\ref{fig:params} shows the observed frequency shifts as a function of Rossby number Ro normalized to the solar value. Ro is the ratio between the rotation period and the convective turnover timescale. The convective turnover timescale $\tau_\text{c}$ is computed following the procedure described in \citet{vanSaders2016}, but with longer chains. Having only the Rossby numbers for 24 stars in the sample with rotation estimate, there is no clear relation between frequency shifts and Rossby number. The Spearman's correlation coefficient is $\sim-0.11$, which confirms a weak negative correlation. A negative correlation between magnetic activity and Rossby number is expected \citep[e.g.][]{Noyes1984b,Mamajek2008}. However, due to the reduced number of data points (only 24) no strong conclusions can be made for this relationship.

Table~\ref{tab:Spear} lists the values for the Spearman's correlation coefficient found between the observed frequency shifts and each stellar property.

\begin{table}[h]\centering
\begin{tabular}{r|c}
{\color{white}empty}& max. \dnu\ variation\\\hline\hline
\Rhk & 0.69\\
\sph & 0.13\\\hline
\teff & 0.68\\
Age & -0.66\\
\prot & -0.61\\
\metal\ (all) & 0.09\\
\metal\ (young) & 0.39\\
\metal\ (old) & -0.46\\
Ro & 0.11\\\hline\hline
\end{tabular}\caption{Spearman's correlation coefficient between the observed frequency shifts and the other stellar properties. The first part of the table concern the correlation between frequency shifts and the measurements of chromospheric and photometric activity. See Appendix~\ref{app:cluster} for details on the two groups of young and old stars in the \dnu-\metal\ relation.}\label{tab:Spear}
\end{table}

Appendix~\ref{app:fit} summarizes the results from a linear regression with multiple variables in order to account for the several potential dependencies. However, rotation periods and Rossby numbers are not available for all targets. Therefore, we first fit the observed frequency shifts taking into account the stellar parameters (\teff, $\log\,g$, \metal, age) available for all the stars in the sample. Taking the residuals between the observed frequency shifts and the predicted values from the fit, we then fit the residuals taking into account the rotation period and Rossby number individually. The results from this procedure confirm significant \dnu-\teff\ and \dnu-age relations.

Finally, in the relations shown in Figure \ref{fig:params}, the Sun behaves normally in comparison the solar-type stars in the sample, with its maximum frequency-shift variation being consistent with that of its neighbors (except KIC~8006161).

\section{Conclusions}\label{sec:conclusions}

In this work, we search for relations between the observed variations in the mode frequencies \citep[measured by][]{Santos2018} and stellar properties. The goal was to identify frequency-shift dependencies on the different stellar properties and understand if those are consistent with a magnetic activity origin.

The target sample for this study comprises the 75 highest signal-to-noise solar-type stars analyzed by \citet{Santos2018}. Such a large sample allows for a better characterization of the observed frequency shifts in solar-type stars.

We start by comparing the observed frequency shifts with the chromospheric activity level available only for 30 out of the 75 stars, which is unfortunate. Nevertheless, we found a strong correlation between frequency shifts and chromospheric activity. One expects that the stronger the activity cycle is, the larger the impact on the mode properties will be, in particular frequencies. Thus, the results are consistent with the observed frequency shifts being indeed activity-related. However, we did not find the same relation between frequency shifts and the photometric activity proxy. This may be because the \sph\ index we calculate is only a lower limit of the true photometric activity due to stellar inclination.

Based on the results for the 75 stars, we found that frequency shifts generally increase with stellar effective temperature (frequency shifts and effective temperature are strongly correlated). These results favor the theoretical predictions for activity-related frequency shifts by \citet{Metcalfe2007}. In what concerns the dependency with effective temperature, our results do not support the theoretical predictions by \citet{Chaplin2007a}.

Frequency shifts are found to be strongly anti-correlated with age, i.e. frequency shifts decrease with age. These results are consistent with the predictions by \citet{Chaplin2007a} and \citet{Metcalfe2007} and consistent with a decreasing magnetic activity as stars evolve \citep[e.g.][]{Wilson1963,Wilson1964}.

Frequency shifts are also found to increase as the surface rotation period decreases (strong anti-correlation). This is in agreement with faster rotators being more active than slower rotators \citep[e.g.][]{Vaughan1981,Baliunas1983,Noyes1984b} and one may thus expect a stronger activity-related effect on the mode frequencies for faster rotators. \prot\ estimates are only available for 50 stars. For a small subset of the target sample (24 stars), we did not find a clear relation between frequency shifts and Rossby number (weak anti-correlation).

Although there is no clear relationship between the observed frequency shifts and stellar metallicity for the full sample (we found a very weak correlation), we do find an increase in the frequency shifts with metallicity for the youngest stars (moderate correlation). These results support the analysis by \citet{Karoff2018}, who interpreted the strong activity cycle of KIC~8006161 (also reflected on the frequency shifts) as an outcome of a deep convection zone due to its high metallicity.

Furthermore, there is a clear outlier on the \dnu-\teff\ and \dnu-\prot\ relationships: KIC~8006161. As mentioned before, KIC~8006161 exhibits a significantly stronger activity cycle than the Sun \citep{Karoff2018}.

We note that for some targets in the sample the measured frequency shifts are a lower limit of the activity-related frequency shifts, as we may not have access to a complete active cycle.

Finally, the activity-related frequency shifts for the Sun are consistent with the general results, behaving normally in comparison with the \kep\ solar-type stars analyzed here.

\citet{Kiefer2017} have already searched for relations between frequency shifts and effective temperature, age, and surface rotation. However, the authors were limited by large uncertainties on the frequency shifts and a small target sample (composed of 24 \kep\ stars). Therefore, their results were not very conclusive, in particular for the relation with effective temperature. Nevertheless, for the three stellar properties considered in \citet{Kiefer2017}, the general trends agree with the results presented here. The results we present are based on a sample more than three times larger than that considered in \citet{Kiefer2017}, spanning a wider parameter space, namely higher effective temperatures and shorter rotation periods. This larger sample allows for a better identification of the relationships between frequency shifts and other stellar properties.

This work, which follows the analysis in \citet{Santos2018}, confirms the prospects of using asteroseismology to learn about stellar magnetism.

\acknowledgments

The authors thank W. Dziembowski for his contribution to the theoretical predictions for activity-related frequency shifts in \citet{Metcalfe2007}. The authors thank D. Bossini for providing the evolutionary tracks shown in this paper.
ARGS acknowledges the support from National Aeronautics and Space Administration under Grant NNX17AF27G. TLC acknowledges support from the European Union’s Horizon 2020 research and innovation programme under the Marie Sk\l{}odowska-Curie grant agreement No.~792848 and from grant CIAAUP-12/2018-BPD. WJC, GRD, RH, and YE acknowledge the support of the UK Science and Technology Facilities Council (STFC). MSC acknowledges support from FCT - Funda\c{c}\~{a}o para a Ci\^{e}ncia e a Tecnologia - through national funds and by FEDER through COMPETE2020 - Programa Operacional Competitividade e Internacionaliza\c{c}\~{a}o - through the grants: UID/FIS/04434/2019; PTDC/FIS-AST/30389/2017 \& POCI-01-0145-FEDER-030389. CK is supported by the Villum Foundation. Funding for the Stellar Astrophysics Centre is provided by the Danish National Research Foundation (Grant agreement No.: DNRF106). TSM acknowledges support from a Visiting Fellowship at the Max Planck Institute for Solar System Research. SM acknowledges the support from the Ramon y Cajal fellowship number RYC-2015-17697. RAG acknowledge the support from PLATO and GOLF CNES grants. MNL acknowledges the support of The Danish Council for Independent Research | Natural Science (Grant DFF-4181-00415). RK acknowledges the support of the STFC consolidated grant ST/P000320/1. VSA acknowledges support from the Independent Research Fund Denmark (Research grant 7027-00096B).

\software{SciPy \citep{SciPy}, NumPy \citep{Numpy}, Matplotlib \citep{matplotlib}, scikit-learn \citep{sklearn}, MESA \citep{Paxton2011,Paxton2013,Paxton2015,Paxton2018}}

\bibliographystyle{aasjournal}
\bibliography{peak-bagging} 

\begin{thebibliography}{}
\expandafter\ifx\csname natexlab\endcsname\relax\def\natexlab#1{#1}\fi
\providecommand{\url}[1]{\href{#1}{#1}}

\bibitem[{Appourchaux {et~al.}(2012)Appourchaux, Benomar, Gruberbauer, Chaplin,
  Garc{\'i}a, Handberg, Verner, Antia, Campante, Davies, Deheuvels, Hekker,
  Howe, Salabert, Bedding, White, Houdek, Silva~Aguirre, Elsworth, {van Cleve},
  Clarke, Hall, \& Kjeldsen}]{Appourchaux2012}
Appourchaux, T., Benomar, O., Gruberbauer, M., {et~al.} 2012, A\&A, 537, A134

\bibitem[{Baglin {et~al.}(2006)Baglin, Michel, Auvergne, \& {COROT
  Team}}]{Baglin2006}
Baglin, A., Michel, E., Auvergne, M., \& {COROT Team}. 2006, in Proceedings of
  {{SOHO}} 18/{{GONG}} 2006/{{HELAS I}}, {{Beyond}} the Spherical {{Sun}}, Vol.
  624, 34.1

\bibitem[{Baliunas {et~al.}(1983)Baliunas, Hartmann, Noyes, Vaughan, Preston,
  Frazer, Lanning, Middelkoop, \& Mihalas}]{Baliunas1983}
Baliunas, S.~L., Hartmann, L., Noyes, R.~W., {et~al.} 1983, ApJ, 275, 752

\bibitem[{Barnes(2007)}]{Barnes2007}
Barnes, S.~A. 2007, ApJ, 669, 1167

\bibitem[{Borucki {et~al.}(2010)Borucki, Koch, Basri, Batalha, Brown, Caldwell,
  Caldwell, {Christensen-Dalsgaard}, Cochran, DeVore, Dunham, Dupree, Gautier,
  Geary, Gilliland, Gould, Howell, Jenkins, Kondo, Latham, Marcy, Meibom,
  Kjeldsen, Lissauer, Monet, Morrison, Sasselov, Tarter, Boss, Brownlee, Owen,
  Buzasi, Charbonneau, Doyle, Fortney, Ford, Holman, Seager, Steffen, Welsh,
  Rowe, Anderson, Buchhave, Ciardi, Walkowicz, Sherry, Horch, Isaacson,
  Everett, Fischer, Torres, Johnson, Endl, MacQueen, Bryson, Dotson, Haas,
  Kolodziejczak, Van~Cleve, Chandrasekaran, Twicken, Quintana, Clarke, Allen,
  Li, Wu, Tenenbaum, Verner, Bruhweiler, Barnes, \& Prsa}]{Borucki2010}
Borucki, W.~J., Koch, D., Basri, G., {et~al.} 2010, Science, 327, 977

\bibitem[{Brun \& Browning(2017)}]{Brun2017}
Brun, A.~S., \& Browning, M.~K. 2017, Living Rev. Solar Phys., 14, 4

\bibitem[{Chaplin {et~al.}(2007)Chaplin, Elsworth, Houdek, \&
  New}]{Chaplin2007a}
Chaplin, W.~J., Elsworth, Y., Houdek, G., \& New, R. 2007, MNRAS, 377, 17

\bibitem[{Chaplin {et~al.}(1998)Chaplin, Elsworth, Isaak, Lines, McLeod,
  Miller, \& New}]{Chaplin1998}
Chaplin, W.~J., Elsworth, Y., Isaak, G.~R., {et~al.} 1998, MNRAS, 300, 1077

\bibitem[{Davies {et~al.}(2015)Davies, Chaplin, Farr, Garc{\'i}a, Lund, Mathis,
  Metcalfe, Appourchaux, Basu, Benomar, Campante, Ceillier, Elsworth, Handberg,
  Salabert, \& Stello}]{Davies2015}
Davies, G.~R., Chaplin, W.~J., Farr, W.~M., {et~al.} 2015, MNRAS, 446, 2959

\bibitem[{Davies {et~al.}(2016)Davies, Silva~Aguirre, Bedding, Handberg, Lund,
  Chaplin, Huber, White, Benomar, Hekker, Basu, Campante,
  {Christensen-Dalsgaard}, Elsworth, Karoff, Kjeldsen, Lundkvist, Metcalfe, \&
  Stello}]{Davies2016}
Davies, G.~R., Silva~Aguirre, V., Bedding, T.~R., {et~al.} 2016, MNRAS, 456,
  2183

\bibitem[{Dziembowski(2007)}]{Dziembowski2007}
Dziembowski, W.~A. 2007, in Unsolved {{Problems}} in {{Stellar Physics}}: {{A
  Conference}} in {{Honor}} of {{Douglas Gough}}, Vol. 948, {eprint:
  arXiv:0709.2602}, 287--294

\bibitem[{Elsworth {et~al.}(1990)Elsworth, Howe, Isaak, McLeod, \&
  New}]{Elsworth1990}
Elsworth, Y., Howe, R., Isaak, G.~R., McLeod, C.~P., \& New, R. 1990, Nat.,
  345, 322

\bibitem[{Fr{\"o}hlich {et~al.}(1995)Fr{\"o}hlich, Romero, Roth, Wehrli,
  Andersen, Appourchaux, Domingo, Telljohann, Berthomieu, Delache, Provost,
  Toutain, Crommelynck, Chevalier, Fichot, D{\"a}ppen, Gough, Hoeksema,
  Jim{\'e}nez, G{\'o}mez, Herreros, Cort{\'e}s, Jones, Pap, \&
  Willson}]{Frohlich1995}
Fr{\"o}hlich, C., Romero, J., Roth, H., {et~al.} 1995, Sol. Phys., 162, 101

\bibitem[{Fr{\"o}hlich {et~al.}(1997)Fr{\"o}hlich, Andersen, Appourchaux,
  Berthomieu, Crommelynck, Domingo, Fichot, Finsterle, Gomez, Gough, Jimenez,
  Leifsen, Lombaerts, Pap, Provost, Cortes, Romero, Roth, Sekii, Telljohann,
  Toutain, \& Wehrli}]{Frohlich1997}
Fr{\"o}hlich, C., Andersen, B.~N., Appourchaux, T., {et~al.} 1997, Sol. Phys.,
  170, 1

\bibitem[{Garc{\'i}a {et~al.}(2010)Garc{\'i}a, Mathur, Salabert, Ballot,
  Regulo, Metcalfe, \& Baglin}]{Garcia2010}
Garc{\'i}a, R.~A., Mathur, S., Salabert, D., {et~al.} 2010, Science, 329, 1032

\bibitem[{Garc{\'i}a {et~al.}(2014)Garc{\'i}a, Ceillier, Salabert, Mathur, {van
  Saders}, Pinsonneault, Ballot, Beck, Bloemen, Campante, Davies, {do
  Nascimento}, Mathis, Metcalfe, Nielsen, Su{\'a}rez, Chaplin, Jim{\'e}nez, \&
  Karoff}]{Garcia2014}
Garc{\'i}a, R.~A., Ceillier, T., Salabert, D., {et~al.} 2014, A\&A, 572, A34

\bibitem[{Goldreich \& Keeley(1977)}]{Goldreich1977}
Goldreich, P., \& Keeley, D.~A. 1977, ApJ, 212, 243

\bibitem[{Hartmann {et~al.}(1984)Hartmann, Soderblom, Noyes, Burnham, \&
  Vaughan}]{Hartmann1984}
Hartmann, L., Soderblom, D.~R., Noyes, R.~W., Burnham, N., \& Vaughan, A.~H.
  1984, ApJ, 276, 254

\bibitem[{Howe {et~al.}(2015)Howe, Davies, Chaplin, Elsworth, \&
  Hale}]{Howe2015}
Howe, R., Davies, G.~R., Chaplin, W.~J., Elsworth, Y.~P., \& Hale, S.~J. 2015,
  MNRAS, 454, 4120

\bibitem[{Hunter(2007)}]{matplotlib}
Hunter, J.~D. 2007, Computing in Science \& Engineering, 9, 90

\bibitem[{Isaacson \& Fischer(2010)}]{Isaacson2010}
Isaacson, H., \& Fischer, D. 2010, ApJ, 725, 875

\bibitem[{Jim{\'e}nez {et~al.}(2002)Jim{\'e}nez, Roca~Cort{\'e}s, \&
  {Jim{\'e}nez-Reyes}}]{Jimenez2002}
Jim{\'e}nez, A., Roca~Cort{\'e}s, T., \& {Jim{\'e}nez-Reyes}, S.~J. 2002, Sol.
  Phys., 209, 247

\bibitem[{Jones {et~al.}(2001--)Jones, Oliphant, Peterson, {et~al.}}]{SciPy}
Jones, E., Oliphant, T., Peterson, P., {et~al.} 2001--, {SciPy}: Open source
  scientific tools for {Python}, ,

\bibitem[{Karoff {et~al.}(2009)Karoff, Metcalfe, Chaplin, Elsworth, Kjeldsen,
  Arentoft, \& Buzasi}]{Karoff2009}
Karoff, C., Metcalfe, T.~S., Chaplin, W.~J., {et~al.} 2009, MNRAS, 399, 914

\bibitem[{Karoff {et~al.}(2013)Karoff, Metcalfe, Chaplin, Frandsen, Grundahl,
  Kjeldsen, {Christensen-Dalsgaard}, Nielsen, Frimann, Thygesen, Arentoft,
  Amby, Sousa, \& Buzasi}]{Karoff2013a}
---. 2013, MNRAS, 433, 3227

\bibitem[{Karoff {et~al.}(2018)Karoff, Metcalfe, Santos, Montet, Isaacson,
  Witzke, Shapiro, Mathur, Davies, Lund, Garcia, Brun, Salabert, Avelino, {van
  Saders}, Egeland, Cunha, Campante, Chaplin, Krivova, Solanki, Stritzinger, \&
  Knudsen}]{Karoff2018}
Karoff, C., Metcalfe, T.~S., Santos, {\^A}. R.~G., {et~al.} 2018, ApJ, 852, 46

\bibitem[{Kiefer {et~al.}(2017)Kiefer, Schad, Davies, \& Roth}]{Kiefer2017}
Kiefer, R., Schad, A., Davies, G., \& Roth, M. 2017, A\&A, 598, A77

\bibitem[{Libbrecht \& Woodard(1990)}]{Libbrecht1990a}
Libbrecht, K.~G., \& Woodard, M.~F. 1990, Nat., 345, 779

\bibitem[{Lund {et~al.}(2017)Lund, Silva~Aguirre, Davies, Chaplin,
  {Christensen-Dalsgaard}, Houdek, White, Bedding, Ball, Huber, Antia,
  Lebreton, Latham, Handberg, Verma, Basu, Casagrande, Justesen, Kjeldsen, \&
  Mosumgaard}]{Lund2017}
Lund, M.~N., Silva~Aguirre, V., Davies, G.~R., {et~al.} 2017, ApJ, 835, 172

\bibitem[{Mamajek \& Hillenbrand(2008)}]{Mamajek2008}
Mamajek, E.~E., \& Hillenbrand, L.~A. 2008, ApJ, 687, 1264

\bibitem[{Marcy {et~al.}(2014)Marcy, Isaacson, Howard, Rowe, Jenkins, Bryson,
  Latham, Howell, Thomas N.~Gautier, Batalha, Rogers, {David Ciardi}, Fischer,
  Gilliland, Kjeldsen, {Christensen-Dalsgaard}, {Daniel Huber}, Chaplin, Basu,
  Buchhave, Quinn, Borucki, Koch, Hunter, Caldwell, Cleve, Kolbl, Weiss,
  Petigura, {Sara Seager}, Morton, Johnson, Ballard, Burke, Cochran, {Michael
  Endl}, MacQueen, Everett, Lissauer, Ford, Torres, {Francois Fressin}, Brown,
  Steffen, Charbonneau, Basri, Sasselov, Winn, {Sanchis-Ojeda}, Christiansen,
  Adams, {Christopher Henze}, Dupree, Fabrycky, Fortney, Tarter, Holman, {Peter
  Tenenbaum}, Shporer, Lucas, Welsh, Orosz, Bedding, Campante, Davies,
  Elsworth, Handberg, Hekker, Karoff, Kawaler, Lund, {M. Lundkvist}, Metcalfe,
  Miglio, Aguirre, Stello, White, Boss, Devore, {Alan Gould}, Prsa, Agol,
  Barclay, Coughlin, Brugamyer, Mullally, Quintana, Still, Thompson, Morrison,
  Twicken, D{\'e}sert, {Josh Carter}, Crepp, H{\'e}brard, Santerne, Moutou,
  Sobeck, {Douglas Hudgins}, Haas, Robertson, {Lillo-Box}, \&
  Barrado}]{Marcy2014}
Marcy, G.~W., Isaacson, H., Howard, A.~W., {et~al.} 2014, ApJS, 210, 20

\bibitem[{Mathur {et~al.}(2014)Mathur, Garc{\'i}a, Ballot, Ceillier, Salabert,
  Metcalfe, R{\'e}gulo, Jim{\'e}nez, \& Bloemen}]{Mathur2014}
Mathur, S., Garc{\'i}a, R.~A., Ballot, J., {et~al.} 2014, A\&A, 562, A124

\bibitem[{Metcalfe {et~al.}(2007)Metcalfe, Dziembowski, Judge, \&
  Snow}]{Metcalfe2007}
Metcalfe, T.~S., Dziembowski, W.~A., Judge, P.~G., \& Snow, M. 2007, MNRAS,
  379, L16

\bibitem[{Middelkoop(1982)}]{Middlekoop1982}
Middelkoop, F. 1982, A\&A, 107, 31

\bibitem[{Noyes {et~al.}(1984)Noyes, Hartmann, Baliunas, Duncan, \&
  Vaughan}]{Noyes1984b}
Noyes, R.~W., Hartmann, L.~W., Baliunas, S.~L., Duncan, D.~K., \& Vaughan,
  A.~H. 1984, ApJ, 279, 763

\bibitem[{Paxton {et~al.}(2011)Paxton, Bildsten, Dotter, Herwig, Lesaffre, \&
  Timmes}]{Paxton2011}
Paxton, B., Bildsten, L., Dotter, A., {et~al.} 2011, ApJS, 192, 3

\bibitem[{Paxton {et~al.}(2013)Paxton, Cantiello, Arras, Bildsten, Brown,
  Dotter, Mankovich, Montgomery, Stello, Timmes, \& Townsend}]{Paxton2013}
Paxton, B., Cantiello, M., Arras, P., {et~al.} 2013, ApJS, 208, 4

\bibitem[{Paxton {et~al.}(2015)Paxton, Marchant, Schwab, Bauer, Bildsten,
  Cantiello, Dessart, Farmer, Hu, Langer, Townsend, Townsley, \&
  Timmes}]{Paxton2015}
Paxton, B., Marchant, P., Schwab, J., {et~al.} 2015, ApJS, 220, 15

\bibitem[{Paxton {et~al.}(2018)Paxton, Schwab, Bauer, Bildsten, Blinnikov,
  Duffell, Farmer, Goldberg, Marchant, Sorokina, Thoul, Townsend, \&
  Timmes}]{Paxton2018}
Paxton, B., Schwab, J., Bauer, E.~B., {et~al.} 2018, ApJS, 234, 34

\bibitem[{Pedregosa {et~al.}(2011)Pedregosa, Varoquaux, Gramfort, Michel,
  Thirion, Grisel, Blondel, Prettenhofer, Weiss, Dubourg, Vanderplas, Passos,
  Cournapeau, Brucher, Perrot, \& Duchesnay}]{sklearn}
Pedregosa, F., Varoquaux, G., Gramfort, A., {et~al.} 2011, JMLR, 12, 2825

\bibitem[{R{\'e}gulo {et~al.}(2016)R{\'e}gulo, Garc{\'i}a, \&
  Ballot}]{Regulo2016}
R{\'e}gulo, C., Garc{\'i}a, R.~A., \& Ballot, J. 2016, A\&A, 589, A103

\bibitem[{Rutten(1984)}]{Rutten1984}
Rutten, R. G.~M. 1984, A\&A, 130, 353

\bibitem[{Salabert {et~al.}(2017)Salabert, Garc{\'i}a, Jim{\'e}nez, Bertello,
  Corsaro, \& Pall{\'e}}]{Salabert2017}
Salabert, D., Garc{\'i}a, R.~A., Jim{\'e}nez, A., {et~al.} 2017, A\&A, 608, A87

\bibitem[{Salabert {et~al.}(2018)Salabert, R{\'e}gulo, P{\'e}rez~Hern{\'a}ndez,
  \& Garc{\'i}a}]{Salabert2018}
Salabert, D., R{\'e}gulo, C., P{\'e}rez~Hern{\'a}ndez, F., \& Garc{\'i}a, R.~A.
  2018, A\&A, 611, A84

\bibitem[{Salabert {et~al.}(2016{\natexlab{a}})Salabert, R{\'e}gulo,
  Garc{\'i}a, Beck, Ballot, Creevey, P{\'e}rez~Hern{\'a}ndez, {do Nascimento},
  Corsaro, Egeland, Mathur, Metcalfe, Bigot, Ceillier, \&
  Pall{\'e}}]{Salabert2016}
Salabert, D., R{\'e}gulo, C., Garc{\'i}a, R.~A., {et~al.} 2016{\natexlab{a}},
  A\&A, 589, A118

\bibitem[{Salabert {et~al.}(2016{\natexlab{b}})Salabert, Garc{\'i}a, Beck,
  Egeland, Pall{\'e}, Mathur, Metcalfe, {do Nascimento}, Ceillier, Andersen, \&
  Trivi{\~n}o~Hage}]{Salabert2016a}
Salabert, D., Garc{\'i}a, R.~A., Beck, P.~G., {et~al.} 2016{\natexlab{b}},
  A\&A, 596, A31

\bibitem[{Santos {et~al.}(2018)Santos, Campante, Chaplin, Cunha, Lund, Kiefer,
  Salabert, Garc{\'i}a, Davies, Elsworth, \& Howe}]{Santos2018}
Santos, A. R.~G., Campante, T.~L., Chaplin, W.~J., {et~al.} 2018, ApJS, 237, 17

\bibitem[{Silva~Aguirre {et~al.}(2015)Silva~Aguirre, Davies, Basu,
  {Christensen-Dalsgaard}, Creevey, Metcalfe, Bedding, Casagrande, Handberg,
  Lund, Nissen, Chaplin, Huber, Serenelli, Stello, Van~Eylen, Campante,
  Elsworth, Gilliland, Hekker, Karoff, Kawaler, Kjeldsen, \&
  Lundkvist}]{SilvaAguirre2015}
Silva~Aguirre, V., Davies, G.~R., Basu, S., {et~al.} 2015, MNRAS, 452, 2127

\bibitem[{Silva~Aguirre {et~al.}(2017)Silva~Aguirre, Lund, Antia, Ball, Basu,
  {Christensen-Dalsgaard}, Lebreton, Reese, Verma, Casagrande, Justesen,
  Mosumgaard, Chaplin, Bedding, Davies, Handberg, Houdek, Huber, Kjeldsen,
  Latham, White, Coelho, Miglio, \& Rendle}]{SilvaAguirre2017}
Silva~Aguirre, V., Lund, M.~N., Antia, H.~M., {et~al.} 2017, ApJ, 835, 173

\bibitem[{Skumanich(1972)}]{Skumanich1972}
Skumanich, A. 1972, ApJ, 171, 565

\bibitem[{van~der Walt {et~al.}(2011)van~der Walt, Colbert, \&
  Varoquaux}]{Numpy}
van~der Walt, S., Colbert, S.~C., \& Varoquaux, G. 2011, CiSE, 13, 22

\bibitem[{{van Saders} {et~al.}(2016){van Saders}, Ceillier, Metcalfe,
  Silva~Aguirre, Pinsonneault, Garc{\'i}a, Mathur, \& Davies}]{vanSaders2016}
{van Saders}, J.~L., Ceillier, T., Metcalfe, T.~S., {et~al.} 2016, Nat., 529,
  181

\bibitem[{Vaughan {et~al.}(1981)Vaughan, Baliunas, Middelkoop, Hartmann,
  Mihalas, Noyes, \& Preston}]{Vaughan1981}
Vaughan, A.~H., Baliunas, S.~L., Middelkoop, F., {et~al.} 1981, ApJ, 250, 276

\bibitem[{Wenger {et~al.}(2000)Wenger, Ochsenbein, Egret, Dubois, Bonnarel,
  Borde, Genova, Jasniewicz, Lalo{\"e}, Lesteven, \& Monier}]{Wenger2000}
Wenger, M., Ochsenbein, F., Egret, D., {et~al.} 2000, Astron. Astrophys. Suppl.
  Ser., 143, 9

\bibitem[{Wilson(1963)}]{Wilson1963}
Wilson, O.~C. 1963, ApJ, 138, 832

\bibitem[{Wilson \& Skumanich(1964)}]{Wilson1964}
Wilson, O.~C., \& Skumanich, A. 1964, ApJ, 140, 1401

\bibitem[{Woodard \& Noyes(1985)}]{Woodard1985}
Woodard, M.~F., \& Noyes, R.~W. 1985, Nat., 318, 449

\bibitem[{Wright {et~al.}(2004)Wright, Marcy, Butler, \& Vogt}]{Wright2004}
Wright, J.~T., Marcy, G.~W., Butler, R.~P., \& Vogt, S.~S. 2004, ApJS, 152, 261

\end{thebibliography}

\appendix
\section{Chromospheric activity}\label{app}

Table \ref{tab} lists the chromospheric activity for the targets with available measurements and the respective source. $B-V$ color index retrieved from SIMBAD database \citep{Wenger2000}. The \Rhk\ values whose source is \citet{Marcy2014} correspond to the values provided by the authors as $S$ index is not provided. For the remainder values, \Rhk\ was calculated from the chromospheric $S$ index, provided by the different authors, following the transformation by \citet{Noyes1984b}, i.e.:
\begin{equation}
R'_\text{HK}=R_\text{HK}-R_\text{phot},
\end{equation}
where
\begin{equation}
R_\text{HK}=1.34\times10^{-4}C_\text{cf}S.
\end{equation}
For $0.3\leq B-V\leq1.6$, the conversion factor $C_\text{cf}$ is given by \citep[see][]{Middlekoop1982,Rutten1984}
\begin{equation}
\log C_\text{cf}=0.25(B-V)^3-1.33(B-V)^2+0.43(B-V)+0.24.
\end{equation}
$R_\text{phot}$ corresponds to the photometric flux \citep{Hartmann1984}
\begin{equation}
\log R_\text{phot}=-4.898+1.918(B-V)^2-2.893(B-V)^3.
\end{equation}

\begin{table}[h]\centering\begin{tabular}{ccccccc}
KIC &  & $B-V$ & & \Rhk & & Reference\\\hline\hline
  1435467 & & 0.43 & & -4.785 & & 1\\
  2837475 & & 0.45 & & -4.761 & & 1\\
  3544595 & & 0.73 & & -4.975 & & 2\\
  3656476 & & 0.80 & & -5.032 & & 3\\
  4914923 & & 0.60 & & -5.148 & & 1\\
  5184732 & & 0.70 & & -5.131 & & 4\\
  6106415 & & 0.55 & & -4.639 & & 4\\
  6116048 & & 0.59 & & -5.017 & & 1\\
  6521045 & & 0.80 & & -5.042 & & 2\\
  6603624 & & 0.78 & & -5.094 & & 1\\
  6933899 & & 0.57 & & -5.016 & & 1\\
  7206837 & & 0.46 & & -4.951 & & 1\\
  7296438 & & 0.68 & & -5.022 & & 3\\
  7680114 & & 0.69 & & -4.908 & & 3\\
  8006161 & & 0.85 & & -5.030 & & 1\\
\end{tabular}\begin{tabular}{ccccccc}
KIC &  & $B-V$ & & \Rhk & & Reference\\\hline\hline
  8379927 & & 0.57 & & -4.820 & & 1\\
  8694723 & & 0.46 & & -4.811 & & 1\\
  9098294 & & 0.57 & & -5.009 & & 1\\
  9139151 & & 0.52 & & -4.899 & & 1\\
  9139163 & & 0.48 & & -4.942 & & 1\\
  9955598 & & 0.72 & & -5.048 & & 2\\
 10454113 & & 0.52 & & -4.829 & & 1\\
 10644253 & & 0.59 & & -4.689 & & 3\\
 10963065 & & 0.51 & & -5.054 & & 2\\
 11253226 & & 0.39 & & -4.615 & & 1\\
 11295426 & & 0.65 & & -5.153 & & 2\\
 12009504 & & 0.55 & & -4.949 & & 1\\
 12069424 & & 0.64 & & -5.117 & & 5\\
 12069449 & & 0.66 & & -5.106 & & 5\\
 12258514 & & 0.55 & & -4.970 & & 1\\
\end{tabular}
\caption{Chromospheric activity index \Rhk. References: (1) \citet{Karoff2013a}; (2) \citet{Marcy2014}; (3) \citet{Salabert2016a}; (4) \citet{Isaacson2010}; (5) \citet{Wright2004}.}
\label{tab}
\end{table}

\section{Cluster analysis: metallicity dependence}\label{app:cluster}

\begin{figure}
\includegraphics[trim=40 10 0 10mm, clip,width=0.49\hsize,angle=3]{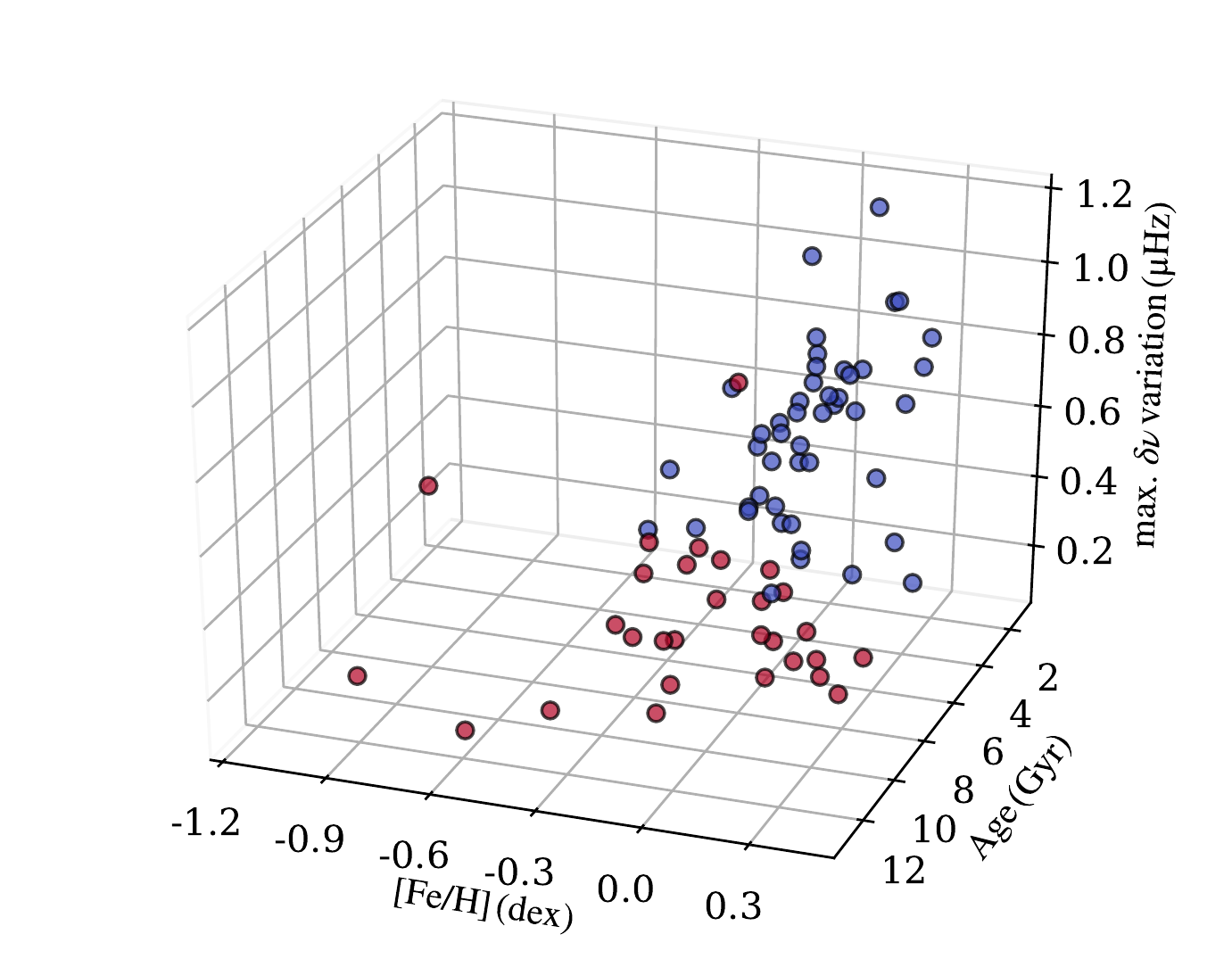}\includegraphics[width=0.5\hsize]{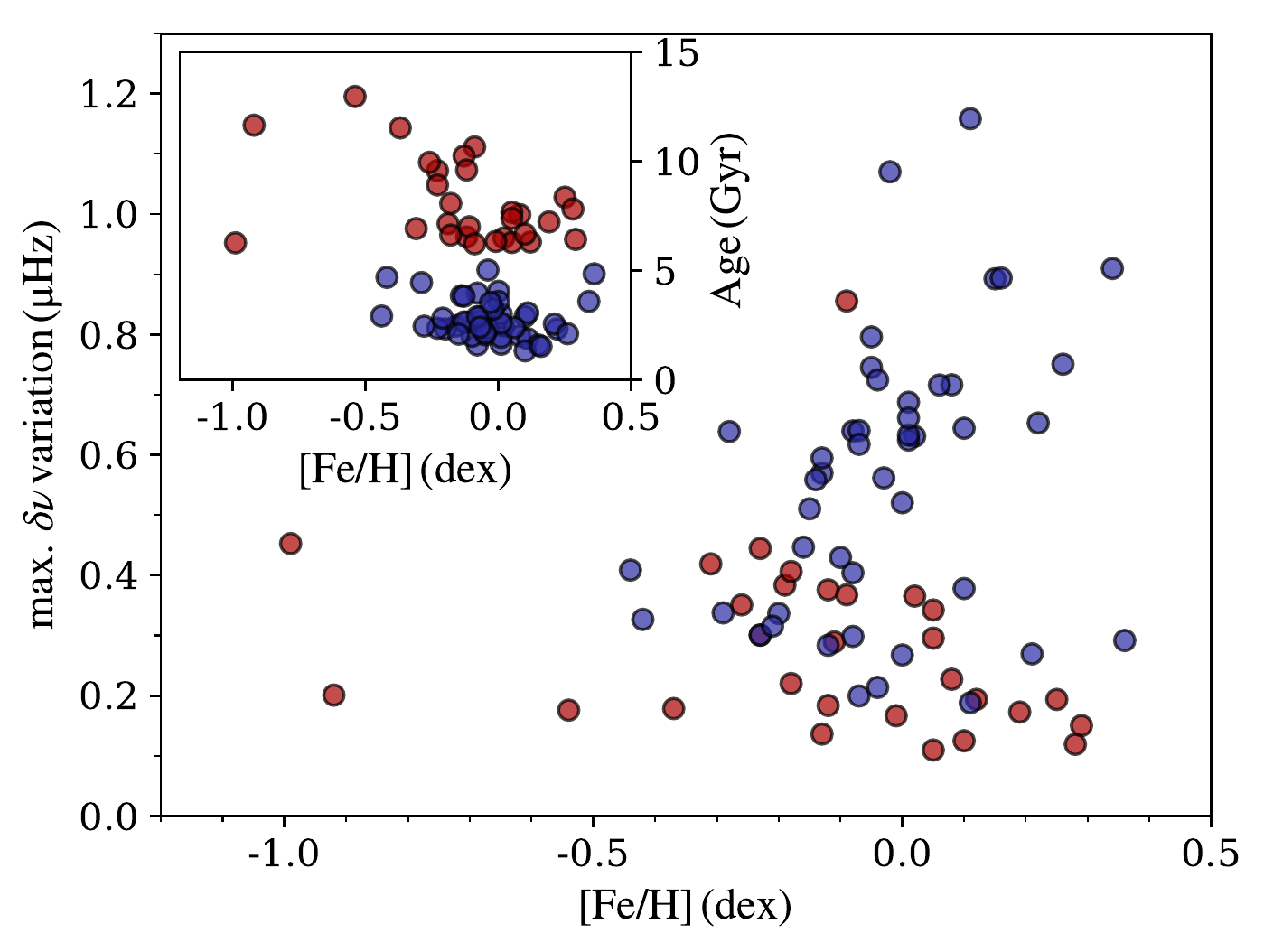}
\caption{{\it Left:} Three-dimensional representation of the \dnu-metallicity diagram (panel f) in Fig.~\ref{fig:params}), where the color code was replaced by the extra dimension. Red and blue symbols show the targets attributed to each cluster. {\it Right:} Same as in the left panel but in two dimensions: observed frequency shifts as a function of metallicity; and stellar age as a function of metallicity.}\label{fig:cluster}
\end{figure}

To better characterize the two sequences seen in the \dnu-metallicity diagram (panel f) in Fig.~\ref{fig:params}) we proceed to a cluster analysis. We use an unsupervised machine learning algorithm to identify the clusters/groups within a three-dimensional dataset: metallicity, age, and observed frequency shifts. The groups are shown in Fig.~\ref{fig:cluster} and their properties are summarized in Table~\ref{tab:cluster}. One of the groups is found to be composed of relatively young stars (younger than 6 Gyr) and shows a moderate correlation between observed frequency shifts and metallicity (see Table~\ref{tab:cluster}). The second group is composed of relatively old stars (older than 6 Gyr) and seems to show the opposite behavior with metallicity (see Table~\ref{tab:cluster}). Frequency shifts are found to decrease with stellar age, thus the average frequency shift is smaller for the group of older stars than for the younger stars. We note again that the observed frequency shifts are likely to be lower limits for some of the targets due to the limited length of observations.

\begin{table}\centering
\begin{tabular}{c|cccc}
& & & &  Spearman's\\
& $\langle$Age$\rangle$ (Gyr) & $\langle\text{[Fe/H]}\rangle$ (dex) & $\langle\delta\nu\rangle$ ($\mu$Hz) & correlation coef.\\\hline\hline
cluster 1 & 2.78 & -0.03 & 0.55 & 0.39\\
cluster 2 & 8.09 & -0.12 & 0.28 & -0.46\\
\end{tabular}
\caption{Average properties of the two groups in the \dnu-metallicity relation.}\label{tab:cluster}
\end{table}

\section{Linear regression with multiple variables}\label{app:fit}

\begin{figure}[h]\centering
\includegraphics[width=\hsize]{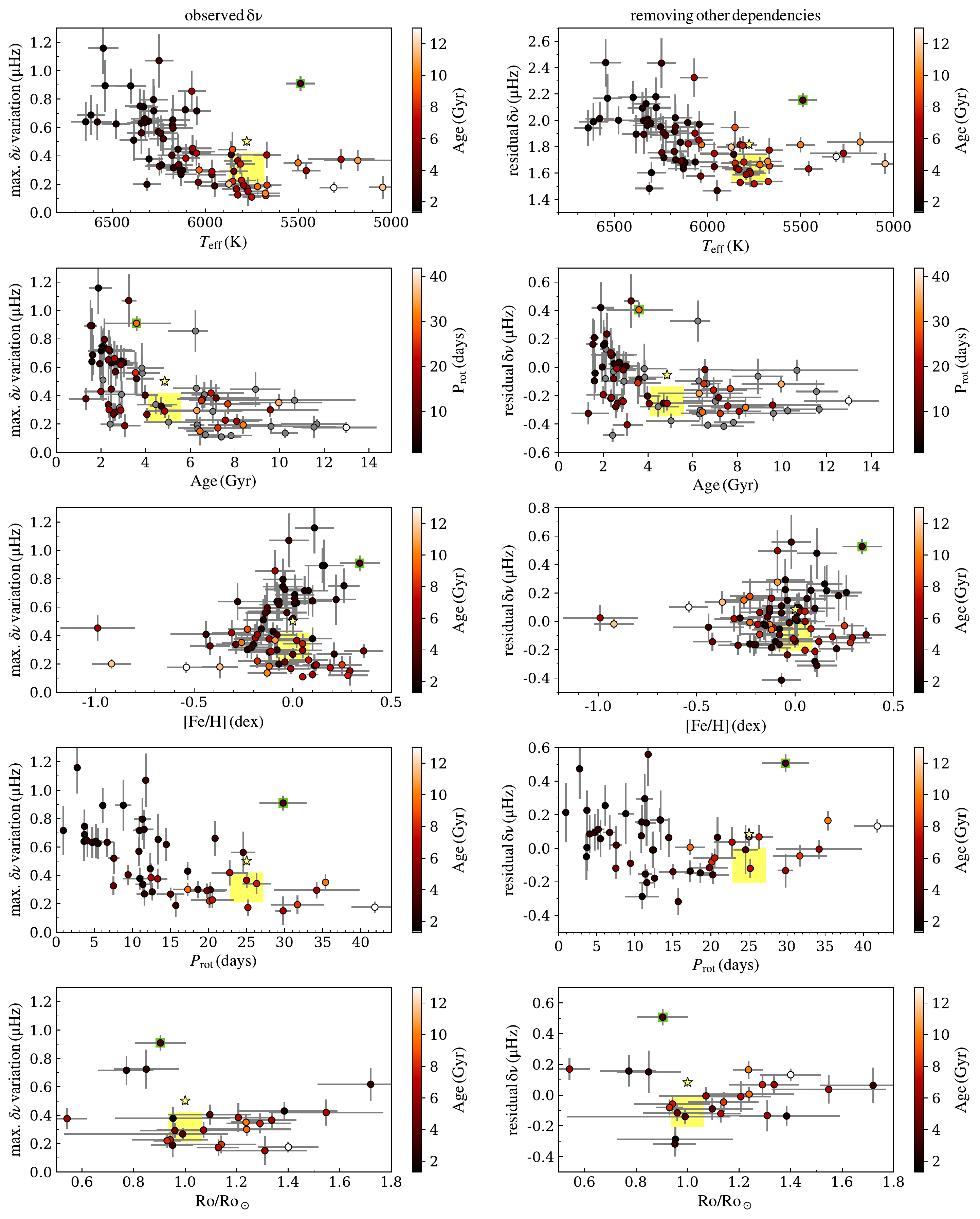}\vspace{-0.3cm}
\caption{{\it Left:} Same frequency-shift relations shown in Fig.~\ref{fig:params}. {\it Right:} Frequency-shift residuals as a function of the different stellar parameters after removing the dependencies on remainder parameters.}\label{fig:regression}\vspace{-0.2cm}
\end{figure}

In order to isolate the frequency-shift dependency on the different parameters, a linear regression with multiple variables is performed. We start by fitting $\delta\nu$ taking into account $T_\text{eff}$, age, metallicity, and $\log\,g$, which are available for the full sample. The first three rows in Fig.~\ref{fig:regression} compares the original observed frequency shifts (left) as a function of \teff, age, and \metal\ with the residuals after subtracting the dependencies on the other parameters (right). Note that in each panel the dependence on the parameter in the horizontal axis is still present. The residuals between the observed frequency shifts and the results from the first fit (with the four parameters) are shown in the two bottom rows. The fitting results are summarized in Table~\ref{tab:regression}. A given variable is considered significant when p-value$<5\%$. Therefore, the regression indicates that only $T_\text{eff}$ and age are significant (for the full sample). The $\delta\nu$-metallicity relation changes significantly after removing the other dependencies, being more compact which results mainly from the age dependency. Nevertheless, the different behavior for young and old stars is still clear. In fact taking the residual frequency shifts shown in Fig.~\ref{fig:regression} and performing the cluster analysis in Appendix~\ref{app:cluster}, we find the same two groups of stars. 

Taking the residuals between the frequency shifts and the original regression, we perform a new fit with rotation period (available for 50 stars). Since rotation also depends on age and \teff, after removing the dependency of \dnu\ on effective temperature and age, the results from the regression suggest that the relation between the residuals and \prot\ is not the significant (fourth row of Fig.~\ref{fig:regression}). We do the same for the Rossby number (available only for 24 stars) and we find the same conclusion (last row).


\begin{table}[h!]\centering
\begin{tabular}{r|cccc}
\multicolumn{5}{c}{fit to max. $\delta\nu$ variation (75 stars)}\\\hline
{\color{white}space}& Coefficient & $\sigma_\text{coef.}$ & p-value & $R^2$ \\\hline
Constant & -2.443 & 1.307 & 0.066 & \multirow{5}{*}{0.408}\\
$T_\text{eff}$ & 0.0003 & 0.0000 & 0.020 & \\
$[\text{Fe}/\text{H}]$ & 0.0586 & 0.109 & 0.591 & \\
$\log\,g$ & 0.285 & 0.176 & 0.110 & \\
Age & -0.028 & 0.014 & 0.048 & \\\hline\hline\multicolumn{5}{c}{{\color{white}empty}}\\
 
\multicolumn{5}{c}{fit to first \dnu\ residuals (50 stars)}\\\hline
{\color{white}space}& Coefficient & $\sigma_\text{coef.}$ & p-value & $R^2$ \\\hline
Constant & -0.3496 & 0.050 &0.000 & \multirow{2}{*}{0.017}\\
$P_\text{rot}$ & -0.003 & 0.003 & 0.368 & \\\hline\hline\multicolumn{5}{c}{{\color{white}empty}}\\

\multicolumn{5}{c}{fit to second \dnu\ residuals(24 stars)}\\\hline
{\color{white}space}& Coefficient & $\sigma_\text{coef.}$ & p-value & $R^2$ \\\hline
Constant & -0.3831 & 0.160 &0.026 & \multirow{2}{*}{0.004}\\
Ro &  -0.0204 & 0.079 & 0.776 & \\\hline\hline\multicolumn{5}{c}{{\color{white}empty}}\\
\end{tabular}
\caption{{\it Top:} Results from the multiple linear regression when fitting the observed frequency shifts with a model that accounts for the global parameters available for the full sample. {\it Middle:} Results from the multiple linear regression when fitting the residual frequency shifts, after removing the dependencies above, taking into account the rotation period. {\it Bottom:} Results from the multiple linear regression when fitting the residual frequency shifts, after removing the dependencies above, taking into account the Rossby number. The table list the respective coefficients and uncertainties, the p-value, and the coefficient of determination $R^2$.}\label{tab:regression}
\end{table}

\end{document}